%% file: JLGomez.tex
\documentclass{evn2004}
\usepackage{txfonts}
\usepackage{graphicx}

\begin{document}

   \title{Real vs. simulated relativistic jets}

   \author{J. L. G\'omez\inst{1,2}, J. M. Mart\'{\i}\inst{3},
          I. Agudo\inst{4}, A. P. Marscher\inst{5}, S. G. Jorstad\inst{5}
          \and
          M. A. Aloy\inst{6}
          }

   \institute{Instituto de Astrof\'{\i}sica de Andaluc\'{\i}a (CSIC), 
              Apartado 3004, E-18080 Granada, Spain
         \and
             Institut d'Estudis Espacials de Catalunya/CSIC, Edif. Nexus,
	     Gran Capita 2--4, E-08034 Barcelona, Spain
	 \and
	     Departamento de Astronom\'{\i}a y Astrof\'{\i}sica, Universidad
	     de Valencia, E-46100 Burjassot (Valencia), Spain
	 \and
	     Max-Planck-Institut f\"ur Radioastronomie, Auf dem H\"ugel, 69,
	     D-53121 Bonn, Germany
         \and
	     Institute for Astrophysical Research, Boston University, 725
	     Commonwealth Avenue, Boston, MA 02215, USA
 	 \and
	     Max-Planck-Institut f\"ur Astrophysik, Karl-Schwarzschild-Str. 1,
	     D-85741 Garching, Germany    
             }

\abstract{

  Intensive VLBI monitoring programs of jets in AGN are showing the existence
of intricate emission patterns, such as upstream motions or slow moving and
quasi-stationary componentes trailing superluminal features. Relativistic
hydrodynamic and emission simulations of jets are in very good agreement with
these observations, proving as a powerful tool for the understanding of the
physical processes taking place in the jets of AGN, microquasars and GRBs.
These simulations show that the variability of the jet emission is the result
of a complex combination of phase motions, viewing angle selection effects,
and non-linear interactions between perturbations and the underlying jet
and/or ambient medium. Both observations and simulations suggest that
shock-in-jet models may be an overly simplistic idealization when
interpreting the emission structure observed in actual jets.
   }

\maketitle

\section{Introduction}

  The improvement in the sensitivity and angular resolution of VLBI
observations has allowed the study of relativistic jets with unprecedented
detail. This advance in the observational techniques has come together with a
rapid development of numerical codes capable of computing the hydrodynamics
of jets with relativistic velocities and energies. Computation of the
emission maps from these models can be used for a direct comparison with
observations, providing therefore a powerful tool for the study of these
objects.

\section{Real jets: observations}

  During the last decade a significant observational effort has
been made to improve our knowledge of the inner jet structure, with special
attention to the magnetic field structure and strength, and its possible
influence in the jet dynamics. Recent observations (Gabuzda 2003) suggests
that the induced transverse magnetic fields are in fact associated with a
toroidal component of the jet magnetic field, instead of reordering by
shocks. This also leaves room for alternative explanations in which jet
components may be associated with kinks in currents originated by the toroidal
fields, instead of strong hydrodynamical shocks (Gabuzda 2003).

  Recent intensive monitoring programs have allowed the study of the inner
jet structure of several sources with the finest time sampling (see e.g.,
G\'omez et al. 2000, 2001; Wehrle et al. 2001; Walker et al. 2001; Marscher et
al. 2002; Stirling et al. 2003; Vermeulen et al. 2003). One of the
best candidates for such intensive monitoring programs is the radio galaxy
3C~120. This is a one of the closest known extragalactic superluminal sources
($z$=0.033) and is a powerful emitter of radiation along the whole
spectrum.

  Monthly 16 epochs of polarimetric 43~GHz VLBA observations of 3C~120
(G\'omez et al. 2001) reveal multiple superluminal components with velocities
in the rage between 4 and 5.8 $h^{-1}_{65} c$. Model fitting of the {\it u-v}
is shown in Fig.~1 ({\it left}). By the end of 1997 the source was observed to
flare, followed by the ejection of a new strong superluminal component, soon
resolved into several distinct features ({\it o1}, {\it o2} and {\it p} in
Fig.~1). These probably do not represent distinct entities but rather
correspond to the complexity of the internal brightness distribution, as shown
in simulations (G\'omez et al. 1997). While subcomponents {\it o1} and {\it
o2} move with a relatively constant velocity, Fig.~1 shows that component {\it
p} splits into two parts that decelerate and decrease in flux more rapidly
than {\it o1} and {\it o2} do. By September 1998 a similar split takes place,
leading to the appearance of components {\it m} and {\it m1}. Two components
closer to the core, labeled {\it r} and {\it s}, are also observable in
Fig.~1. These new components appear in the wake of the main superluminal
feature (containing {\it o1} and {\it o2}) and present significantly slower
(by a factor of $\sim$4) proper motions than any of the other superluminal
components detected in 3C~120. Further evidence for slow moving or
quasi-stationary components trailing superluminal features has been found by
similarly dedicated monitoring programs on other sources (Tingay et al. 2001;
Jorstad et al. 2001).

\begin{figure*}
  \centering
  \includegraphics[width=\textwidth]{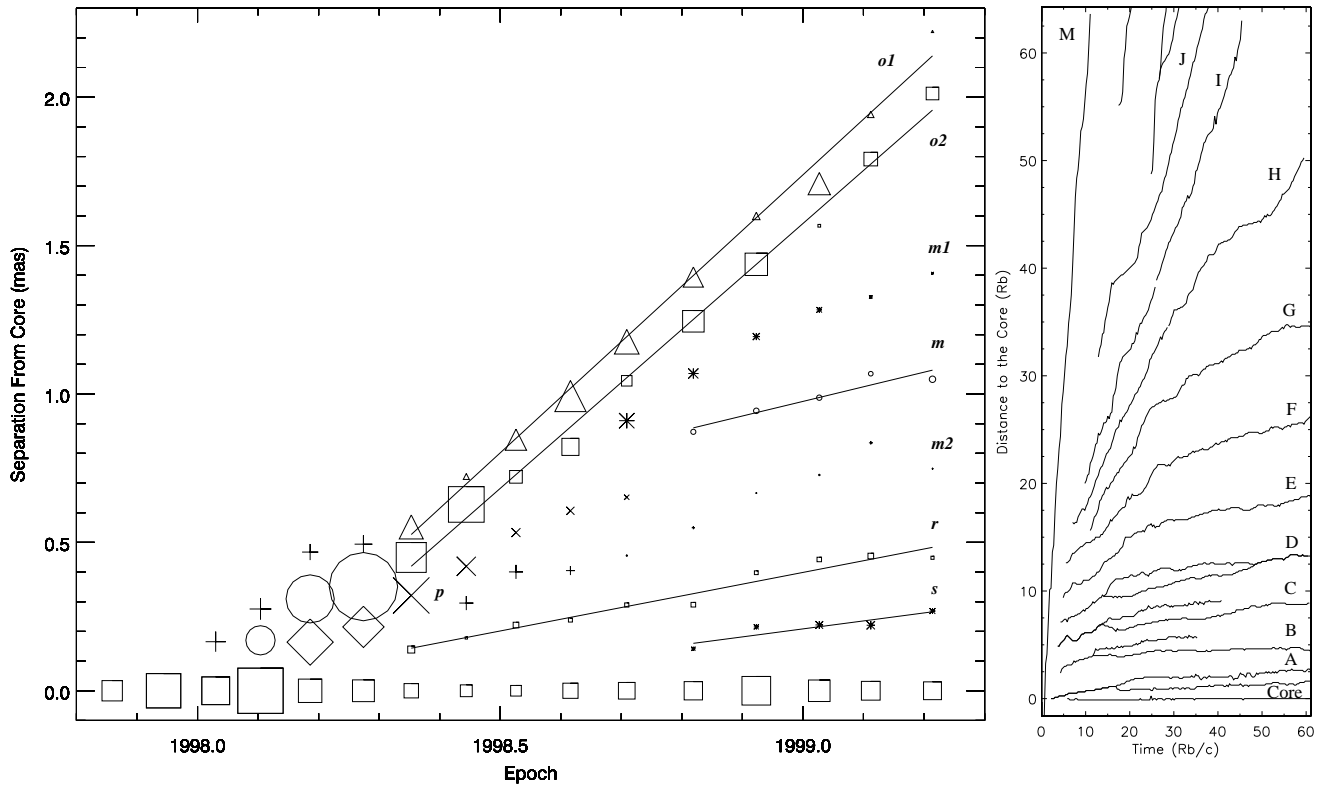}
  \caption{{\it Left}: Projected angular distance from the core vs. time
for the inner jet features of the jet in 3C~120. The symbol size is
proportional to the component's total flux density. Reproduced from G\'omez
et al. (2001). {\it Right}: Distance to the core vs. time for the components
appearing in a relativistic hydrodynamic and emission simulation (Agudo et
al. 2001) of a jet in which a perturbation, associated with the component
labeled {\it M}, has been injected.}
  \label{Trailings}
\end{figure*}

  These intensive monitoring program on 3C~120, consisting also of
simultaneous observations at 22~GHz, led G\'omez et al. (2000) to find a
region in the jet of 3C~120 in which superluminal components present variations in
the total and polarized flux densities with time scales of months,
accompanied by a progressive rotation of the magnetic polarization vector.
This was interpreted as due to the interaction of the jet with the
external medium or a cloud with properties intermediated of those of the broad
and narrow emission-line regions (G\'omez et al. 2000). The rotation of the
magnetic vector was interpreted by these authors as produced by Faraday
rotation of the ionized cloud, the level of which was estimated from the
different polarization angles observed at 22 and 43~GHz.

\begin{figure*}
  \centering
  \includegraphics[width=\textwidth]{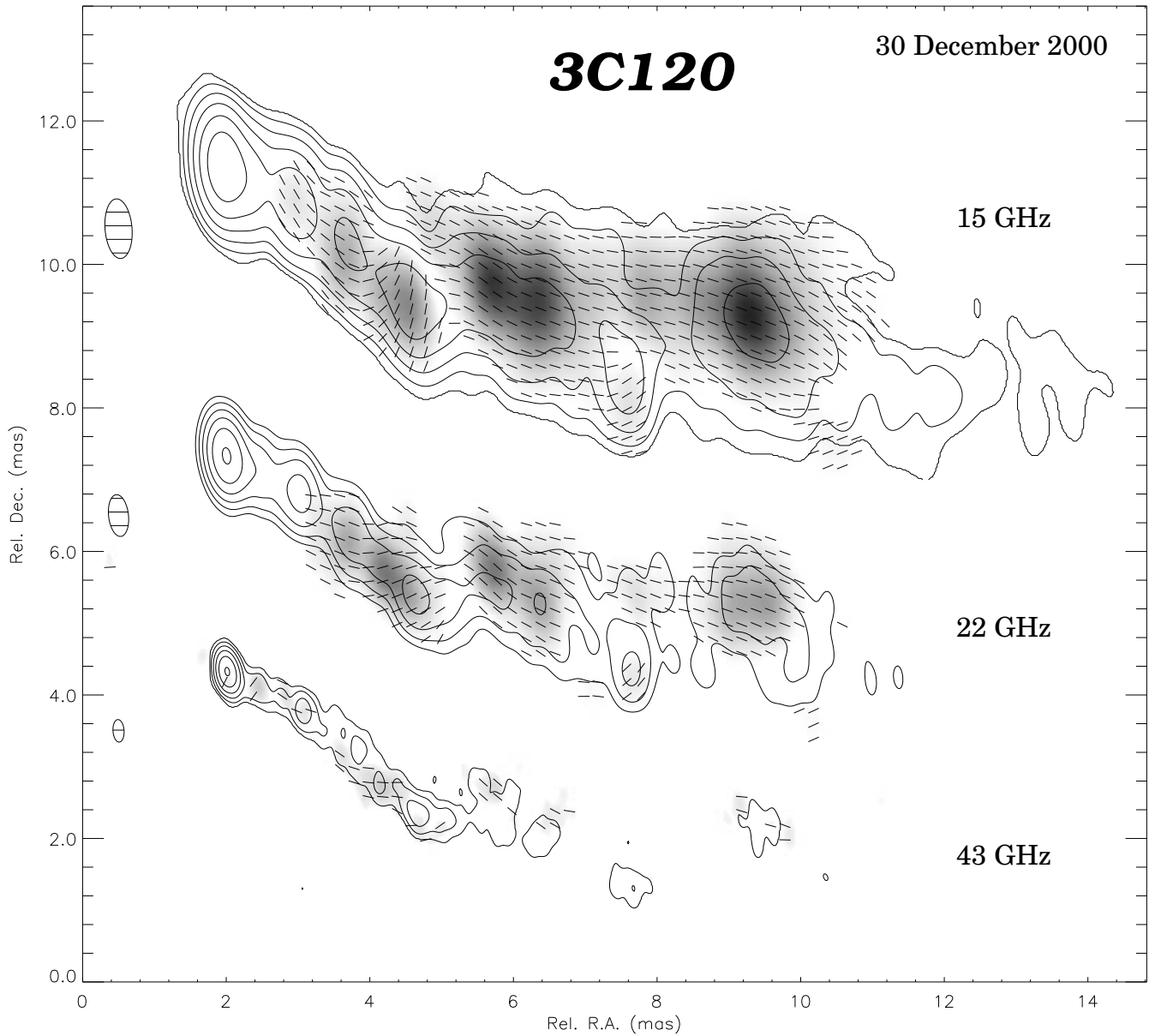}
  \caption{VLBA images of the radio galaxy 3C~120 for 30 December 2000 at 15 
(top), 22 (middle), and 43~GHz (bottom). Total intensity is plotted in
contours at common values for the three images of 0.001 (only for
15~GHz), 0.003, 0.009, 0.026, 0.076, 0.222, and 0.647 Jy/beam. Linear gray
scale shows the linearly polarized intensity. Bars (of unit length) indicate
the direction of the magnetic polarization vector. Convolving beams are
plotted to the left of each image.}
  \label{3C120-Dec00}
\end{figure*}
  
\begin{figure*}
  \centering
  \includegraphics[width=\textwidth]{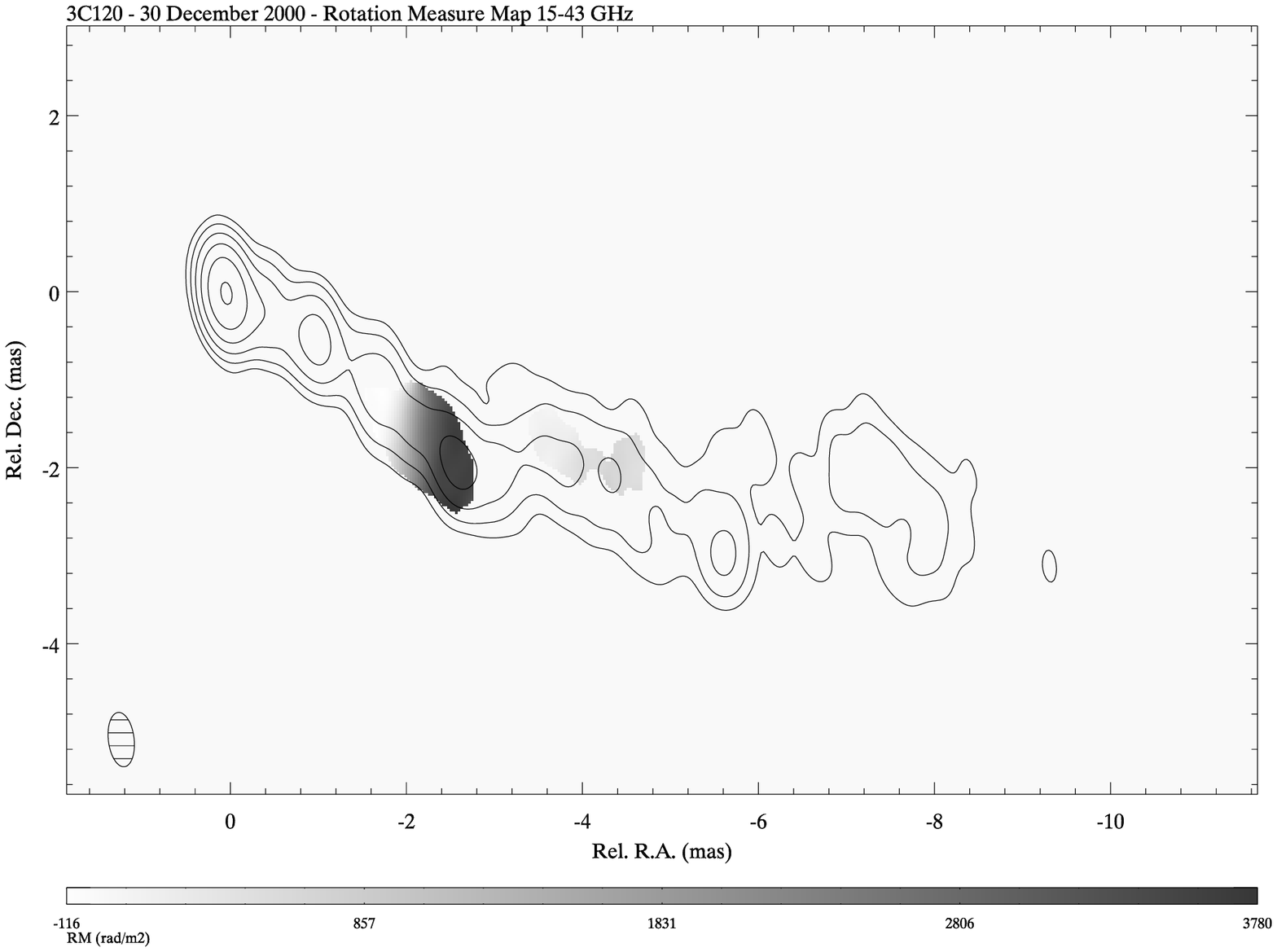}
  \caption{Rotation measure map (gray scale) for the radio galaxy 3C~120 at 
30 December 2000 combining the images shown in Fig.\ref{3C120-Dec00}.
Contours show he total intensity image at 22~GHz.}
  \label{RM}
\end{figure*}

  Later VLBA observations at 15, 22, and 43~GHz (see Figs.~\ref{3C120-Dec00}
and \ref{RM}), confirm the presence of a rotation measure (RM) region at the
same location as that reported by G\'omez et al. (2000). However, the amount
of RM in the region seems to have dropped from $\sim 6000$ to $\sim 3800$ rad
m$^{-2}$ in about two years. This variability may be expected in the case of
being affected by a rapid evolving interaction of the jet with the external
medium or cloud. This Faraday screen is also coincident in position with a
region of increased jet opacity, as shown in Fig.~4 of G\'omez et al. (2000).

  Similar interactions between the jet and external medium have been reported
for other sources. In Gabuzda et al. (2001), the highly bent structure of the
BL~Lac object 0735+178 is interpreted as the interaction of the jet with the
external medium. Increased RM, coincident with a region of enhanced
opacity, is observed at the location where the jet bends by an
angle of about 90$^{\circ}$ in the plane of the sky. In Homan et al. (2003) a
jet interaction with the external medium is considered to explain the
observed deflection of a superluminal component (labeled $C4$) in its motion
along the jet of 3C~279. 

  Most of our knowledge of the nature of relativistic jets comes from the
study of the emission components proper motion and flux evolution.
Superluminal components in jets exhibit ballistic motions away from
the core, as well as curved paths suggestive of streaming motions along a
funnel (e.g., Homan et al. 2001; Lister 2001). Some of these bent jets
resemble helical structures in projection, presumably originated by precession
of the jet nozzle. Growing evidence suggests that this is actually the case
for BL~Lac itself, where components are found to be ejected at different
position angles, initially moving with ballistic trajectories to later on
follow curved paths that are in agreement with a helical jet (Denn, Mutel \&
Marscher 2000; Stirling et al. 2003; Gabuzda \& Cawthorne 2003).

  The radio galaxy 3C~120 also presents significant evidence in favor of a jet
precession, or at least a change in the direction of the jet nozzle. The
images of Fig.\ref{3C120-Dec00}, specially that of the higher resolution
at 43~GHz, shows a jet structure in which the knots of emission plot a
twisted structure resembling that of a helix in projection. This precessing
nature of 3C~120 has also been suggested by studying the changing direction
for the direction of ejection of components (G\'omez et al. 1998).
Further evidence has been found by analyzing the changes in position angle
and magnetic field orientation of the superluminal components in their
motion along the jet, specially in the inner 2 mas from the core (see Fig.~5
of G\'omez et al. 2001).

  Other similar evidence for jet precession is being found in an increasing
number of sources (e.g., Walker et al. 2001; Lister et al. 2003).
Interpretation of this phenomenon, as well as other intricate emission
structure variability such as upstream phase motion of jet features (Wehrle et
al. 2001), requires a detailed modeling of the hydrodynamic and emission
processes taking place in these relativistic sources.

\section{Numerically simulated relativistic jets}

  The development of modern high resolution techniques in numerical
hydrodynamics has allowed the computation of time dependent simulations
of relativistic jets (Mart\'{\i}, M\"uller \& Ib\'a\~nez 1994; Duncan \&
Hughes 1994, and review Mart\'{\i} \& M\"uller 1999). These models are
capable, for the first time, to study the jet dynamics with unprecedented
detail, and under very similar conditions as it is thought are taking place in
real sources (that is, strong shocks, relativistic internal energies and bulk
flow velocities, etc.). Some of the latest simulations have started to explore
three-dimensional relativistic jets (Aloy et al. 2003, and references therein;
Hardee et al. 2001; Hughes, Miller \& Duncan 2002), magnetized relativistic
jets (Komissarov 1999), as well as jet formation and collimation making use of
general relativistic magnetohydrodynamic codes (Koide 2003, and references
therein; Gammie, McKinney \& T\'oth 2003; De Villiers \& Hawley 2003).

  However, the observed emission structure is not just a direct mapping of the
jet hydrodynamical variables (pressure, density, velocity). The final
radiation reaching our detectors is greatly determined by other several
processes, like opacity, particle acceleration, radiative losses, Faraday
rotation, and, most importantly, by relativistic effects such as light
aberration and light travel time delays. For relativistic speeds (and small
viewing angles) time delays can be of such importance as to leave the emission
images with no apparent relationship to the hydrodynamical jet
structure. Hence, the state of the art in the simulation of relativistic
jets involves the computation of the emission, taking into account the
appropriate relativistic and transfer of radiation processes, from the
relativistic hydrodynamic results (G\'omez et al. 1995, 1997; Mioduszewski,
Hughes, \& Duncan 1997; Komissarov \& Falle 1997; Aloy et al. 2000, 2003;
Agudo et al. 2001; and reviews G\'omez 2001, 2002).

  With these new numerical techniques it is now possible to study with great
detail the generation, internal structure, and evolution of strong shock waves
(G\'omez et al. 1997; Mioduszewski, Hughes, \& Duncan 1997; Komissarov \&
Falle 1997). Moving shocks, induced by introducing different type of
perturbations at the jet inlet, provide a good explanation for the overall
properties of superluminal components. In G\'omez et al. (1997) simulations
the propagation of a strong shock through a series of recollimation shocks is
analyzed, showing that the latter may experience a temporary dragging of their
position downstream, followed by upstream motions to recover their initial
locations. These ``wiggling'' of quasi-stationary features do not correspond
to actual fluid motions. On the contrary, they are related to phase motions,
indicating the location of the recollimation shocks which may vary with
changes in the jet hydrodynamic properties (i.e., jet Mach number). As shown
in G\'omez (2002), these phase motions can easily lead to wrong identification
of components when the time sampling of the jet emission structure is not good
enough. Furthermore, measured proper motions may actually depend on the
angular resolution with which the jet is observed, since different
convolving beams will be sensitive to different jet structures.

  Agudo et al. (2001) simulations show that strong jet perturbations (which
can be associated with bright superluminal components) interact with the
underlying jet and external medium as they propagate. This excites pinch-mode
jet-body instabilities, which in turn lead to the formation of recollimation
shocks and rarefactions in the wake of the main perturbation. Figure 1 ({\it
right}) plots the separation from the core as a function of time for these
{\it trailing} components as computed by Agudo et al. (2001). They can be
easily distinguished because they appear to be released from the primary
superluminal component instead of being ejected from the core. The apparent
velocities of the trailing features should range from subluminal closest to
the core, to more superluminal near the leading component.

  Figure 1 provides a one-to-one comparison between the actual inner motions
of components in the radio galaxy 3C~120 and that of the simulations by Agudo
et al. (2001), revealing a very good agreement. We can associate the
new strong superluminal component (containing {\it o1} and {\it o2}; Fig.~1
{\it left}) with the leading perturbation (labeled {\it M} in the
simulations; Fig.~1 {\it right}). Components {\it m1}, {\it m} and {\it m2}
have been observed to emerge in the wake of the main superluminal component,
and to present motions which are significantly smaller (1.2 $h^{-1}_{65} c$
for {\it m}). In addition, components {\it m1}, {\it m}, {\it m2}, {\it r} and
{\it s} present increasing velocities with distance from the core, being the
speeds of {\it r} and {\it s} the smallest (subluminal) detected in the jet of
3C~120.

  Recent three-dimensional simulations have paid special attention to the
response of relativistic jets to precession. Hardee et al. (2001) show that
combination of the helical surface and first-body modes may lead to complex
pressure and velocity structure inside the jet. They appear in synthetic
emission images as differentially moving and stationary features in the jet,
therefore providing an alternative mechanism for the production of jet
components. In Hughes et al. (2002) the analysis of the jet emissivity for a
precessing jet shows that this is in general a complex function of both,
Doppler boosting and jet internal hydrodynamic conditions.

\begin{figure}
  \centering
  \includegraphics[width=8.7cm]{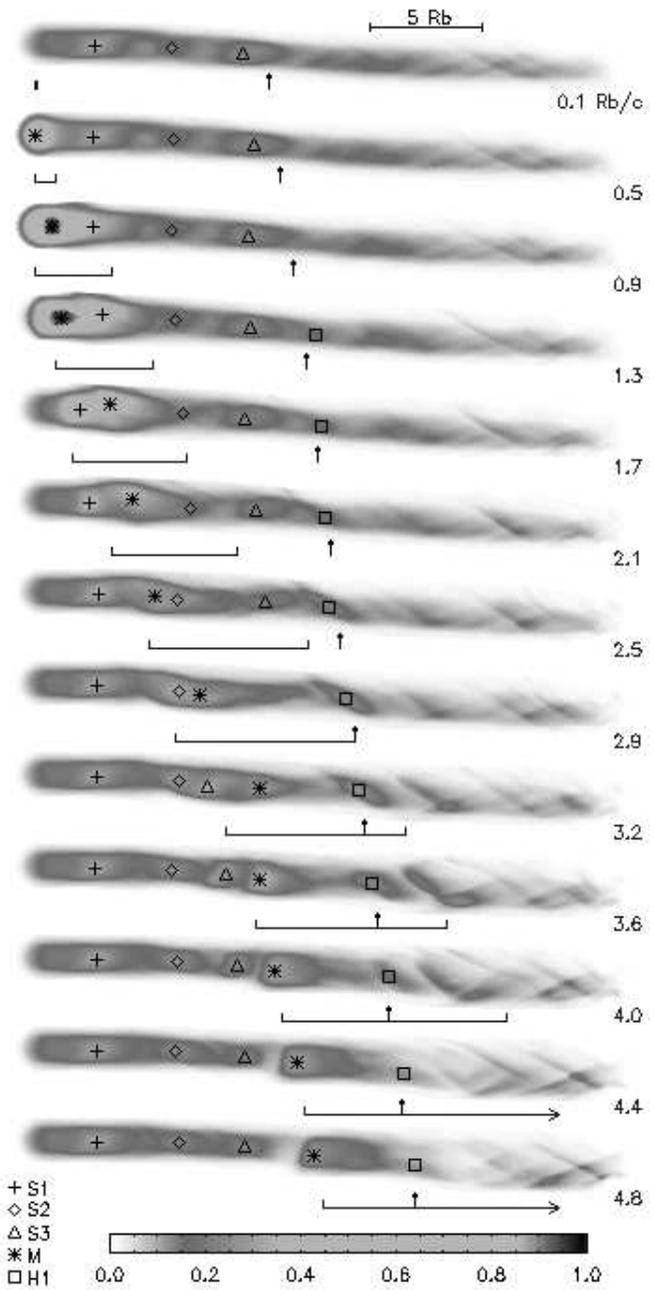}
  \caption{Three-dimensional numerical simulation of a perturbation traveling
along a precessing jet. Images show the time sequence (from top to bottom) of
the computed radio emission (total intensity in arbitrary units) for a
viewing angle of 15$^{\circ}$ and an optically thin frequency. Underbrackets
indicate the extension of the imposed hydrodynamic perturbation. Knots in the
emission are marked with different symbols. Reproduced from Aloy et al.
(2003).}
  \label{3D}
\end{figure}

  First three-dimensional relativistic hydrodynamic and emission
simulations of a precessing jet through which a perturbation (shock wave) is
set to propagate have been carried out by Aloy et al. (2003). Synthetic radio
maps computed from the hydrodynamic model taking into account the appropriate
light travel time delays are shown in Fig.~\ref{3D}. The introduced
perturbation appears in the emission maps as a large region of
enhanced emission. This stretching of the perturbation as seen in the
observer's frame is produced by the light travel time delays between the
front and back of the perturbation. As a consequence, the structure of the
perturbation is magnified leading to brightness distribution variations
within the component as seen in the observer's frame. This could have
significant implications when interpreting the observations of superluminal
sources. Identifying brightness peaks in radio maps with components
may be misleading, as the component in this case may only be one of many
brightness features caused by a single perturbation and may not be related to
any physical structure on its own. Furthermore, a component may arise from
different regions of the perturbation having different hydrodynamical
properties, which could also change with time as the component evolves along
the jet and interacts with the external medium and the underlying jet.

  These simulations are therefore suggesting that shock-in-jet models may be
an overly simplistic idealization when interpreting the emission patterns
observed in actual jets. Indeed, most observable features should not be
related to fluid bulk motions, but instead to a complex combination of bulk
and phase motions, viewing angle selection effects, and non-linear
interactions between perturbations and the external medium and/or underlying
jet.

  The improvement in the numerical modeling of relativistic jets will allow
in the near future the computation of synthetic polarization emission maps
making use of the recently developed relativistic magnetohydrodynamic
simulations. These type of simulations would be capable of exploring the
inner regions of jets in which the magnetic field could be dynamically
important. Other efforts are aimed to the implementation of different
equations of state, to account for the electron energy transport, and the
computation of the synchrotron self Compton emission.

\section{Conclusions}

  Intensive monitoring VLBI programs on multiple jets in AGN are providing
information of the inner emission structure with unprecedented spatial and
temporal resolutions. These are revealing the existence of intricate emission
patterns, such as upstream motions or slow moving and quasi-stationary
components trailing superluminal features. Numerical relativistic hydrodynamic
and emission simulations are in good agreement with the observations,
revealing the importance of such computations for the interpretation of
actual sources. They also show that the non-linear hydrodynamic evolution of
perturbations can determine the observed emission properties so that the
interpretation of observed radio maps is error-prone when naively associating
single shocks to superluminal components.

  Both observations and simulations are suggesting that we are reaching a
level of detail in which our radio images cannot just be idealized as a series
of Gaussian moving components, each associated with a single shock wave. It is
perhaps time to consider new ways to analyze our images, since otherwise we
may overlook very important pieces of information present in our data.

\end{document}